# Transport through a quantum dot subject to spin and charge bias: Kondo regime

R. Świrkowicz[1], J. Barnaś[2,3], M. Wilczyński[1]

[1]Faculty of Physics, Warsaw University of Technology, ul. Koszykowa 75,

00-662 Warsaw, Poland

[2]Department of Physics, Adam Mickiewicz University, ul. Umultowska 85,

61-614 Poznań, Poland

[3]Institute of Molecular Physics, Polish Academy of Sciences, ul. Smoluchowskiego 17,

60-179 Poznań, Poland

**Abstract**

Spin and charge transport through a quantum dot coupled to external nonmagnetic leads is analyzed theoretically in terms of the non-equilibrium Green function formalism based on the equation of motion method. The dot is assumed to be subject to spin and charge bias, and the considerations are focused on the Kondo effect in spin and charge transport. It is shown that the differential spin conductance as a function of spin bias reveals a typical zero-bias Kondo anomaly which becomes split when either magnetic field or charge bias are applied. Significantly different behavior is found for mixed charge/spin conductance. The influence of electron-phonon coupling in the dot on tunneling current as well as on both spin and charge conductance is also analyzed.

Keywords: quantum dots, spin bias, Kondo effect, phonons

PACS: 73.23 Hk; 73.63.Kv; 72.25.-b

# I. INTRODUCTION

There is currently a significant interest in spin and charge transport in mesoscopic systems, as well as in the associated phenomena like spin accumulation, current induced spin dynamics, etc. One of the important issues of this research is related to the problem of spin battery, which could be used as a source of spin current and/or spin voltage. This in turn opens a new field of research, i.e. spin and charge transport in systems subject to spin bias. The main objective of this paper is just to study transport characteristics in the case of quantum dots subject to spin and charge bias.

Electronic transport through quantum dots has been a subject of extensive experimental and theoretical works in recent two decades. Most of efforts have been focused on understanding the Kondo phenomenon in single dots and arrays of dots. The Kondo effect is a many-body phenomenon and is a consequence of spin fluctuations of the dot's spin. These fluctuations lead to a Kondo peak in the density of states at the Fermi level (in an unbiased system), which gives rise to an enhanced conductance in the small bias range (zero-bias peak in conductance).[1-4] It has been shown both theoretically and experimentally, that the Kondo anomaly becomes split and suppressed by an external magnetic field.[5-7] Similar behavior also takes place in the absence of external field, but when the dot is connected to ferromagnetic leads instead of nonmagnetic ones.[8-17]

From a theoretical point of view the Kondo phenomenon has been analyzed by various perturbative and nonperturbative techniques. These include the equation of motion method for calculating the relevant Green functions, real time diagrammatic techniques, poor man scaling approaches, numerical renormalization group method, and others. Some of the techniques, however, describe correctly only certain features of the phenomenon, while the other features may not be properly accounted for. Generally, the results obtained by the numerical renormalization group method are considered as the most reliable ones. However, equation of motion method was widely used to investigate the Kondo anomaly in a variety of nanoscopic systems including molecules and nanotubes (see e.g. Refs 12,13,18-22). Although the method fails to describe properly the symmetric Anderson model with the dot's energy level $E_0$=-$U/2$ at $T$=$0$ $K$, it leads to results qualitatively consistent with other techniques for large values of the correlation parameter $U$, especially at relatively high temperatures $T$~$T_k$.[19,21,22] This method is very useful in nonequilibrium situations, when a finite bias voltage is applied. The method was also used to describe the Kondo physics in quantum dots coupled to ferromagnetic electrodes with collinear and non-collinear magnetic moments giving reasonable results which are consistent with other approaches and also with experimental data.[13,18,19,22]

In a typical experiment the quantum dot is attached to two nonmagnetic leads, in which electrons are assumed to be in thermodynamic equilibrium described by appropriate electrochemical potentials that are independent of spin orientation (charge bias). However, it has been proposed recently that instead of pure charge bias one can also apply a spin bias.[23-29] The latter can be realized experimentally for instance by making use of the phenomenon of spin accumulation at biased contacts between ferromagnetic and nonmagnetic materials, or due to illumination of semiconductor leads with circularly polarized light.[30-34] The electrodes are then not in equilibrium. However, one may assume equilibrium electron distribution in each spin subband separately.

In a recent work Katsura[28] considered a magnetic impurity that was exchange coupled to two external leads, and the analysis was focused on the Kondo effect in electronic transport in the presence of spin dependent bias (charge and spin bias) applied to the system. In this paper we consider a related problem. However, the model assumed in our considerations, as well as the theoretical method we employ, are different from those in Ref. 28. We analyze in detail the interplay of charge and spin voltages applied to a quantum dot and reveal new features shown by spin as well as mixed charge/spin conductance. Very recently a similar problem was considered in Ref. 29. In our paper, however, we include the role of electron-phonon coupling in the dot, which was not taken into account in Refs 28,29.

The paper is organized as follows. In section II we describe the model and outline the theoretical method used to calculate transport properties in the absence of electron-phonon coupling in the dot. The corresponding numerical results are also presented and discussed there. In section III we consider the situation when the electrons in the dot are coupled to a phonon bath. Summary and final conclusions are in section IV.

## II. KONDO EFFECT IN TRANSPORT THROUGH A SINGLE-LEVEL QUANTUM DOT UNDER SPIN AND CHARGE BIAS

In this section we consider electronic transport through a single-level quantum dot in the Kondo regime. The dot is connected to two external nonmagnetic leads, in which the corresponding chemical potentials are generally spin dependent. Such a spin splitting of the chemical potentials can be reached by the methods mentioned in the introduction. We begin with a short description of the model and theoretical method, and then proceed with the description and discussion of the corresponding numerical results.

### A. Model

The system considered in this paper consists of a quantum dot (QD) attached to two external non-magnetic electrodes. In this section we assume that the dot is decoupled from the phonon bath (this assumption will be relaxed in section III). The whole system can be then described by Hamiltonian of the following general form:

$$H = H_L + H_R + H_D + H_T. \qquad (1)$$

The first two terms, $H_\beta = \sum_{\mathbf{k}\sigma} \varepsilon_{\mathbf{k}\beta} c^+_{\mathbf{k}\beta\sigma} c_{\mathbf{k}\beta\sigma}$ for $\beta = L, R$, represent the left (*L*) and right (*R*) electrodes, with $c^+_{\mathbf{k}\beta\sigma}$ ($c_{\mathbf{k}\beta\sigma}$) being the corresponding creation (annihilation) operator of an electron with energy $\varepsilon_{\mathbf{k}\beta}$ and wave vector $\mathbf{k}$. The next term, $H_D$, stands for Hamiltonian of the dot and takes a typical Hubbard form,

$$H_D = \sum_\sigma E_\sigma d^+_\sigma d_\sigma + U d^+_\uparrow d_\uparrow d^+_\downarrow d_\downarrow, \qquad (2)$$

where $E_\sigma$ denotes the dot's discrete energy level, which in general case can be spin dependent, $d^+_\sigma$ and $d_\sigma$ are the relevant creation and annihilation operators of an electron with spin σ (σ = ↑,↓), and *U* stands for the Hubbard electron correlation parameter. The final term, $H_T$, of Hamiltonian (1) describes tunneling processes between the dot and electrodes, and takes the form $H_T = \sum_{\mathbf{k}\beta} \sum_\sigma T_{\mathbf{k}\beta} c^+_{\mathbf{k}\beta\sigma} d_\sigma + h.c.$, where $T_{\mathbf{k}\beta}$ are the tunneling matrix elements. In the following, coupling of the dot to external leads will be described effectively by the parameters $\Gamma_\beta = 2\pi \sum_{\mathbf{k}} |T_{\mathbf{k}\beta}|^2 \delta(E - \varepsilon_{\mathbf{k}\beta})$ (for $\beta = L, R$), which also determine the dot's level width. These parameters are assumed constant within the electron band of the leads and zero otherwise.

Charge current flowing through the system in the spin channel σ is given by the formula[35,36]

$$J_\sigma = \frac{ie}{2\hbar} \int \frac{dE}{2\pi} \left\{ (\Gamma_L - \Gamma_R) G^<_\sigma(E) + [f_{L\sigma}(E)\Gamma_L - f_{R\sigma}(E)\Gamma_R][G^>_\sigma(E) - G^<_\sigma(E)] \right\}, \qquad (3)$$

where $G^<_\sigma(E)$ ($G^>_\sigma(E)$) denotes the Fourier transform of the lesser (greater) Green function, and $f_{\beta\sigma}(E)$ is the Fermi-Dirac distribution function for the $\beta$-th lead. We assume that the electrochemical potentials of the leads are spin dependent, which generally is equivalent to charge $V^c$ and spin $V^s$ bias applied to the system. The electrochemical potentials in the electrodes can be then written in the following form: $\mu_{L\sigma} = \varepsilon_F + \frac{1}{2}e(V^c + \hat{\sigma}V^s)$ and $\mu_{R\sigma} = \varepsilon_F - \frac{1}{2}e(V^c + \hat{\sigma}V^s)$, with $\hat{\sigma} = 1$ ($\hat{\sigma} = -1$) for σ = ↑ (σ = ↓). For simplicity we have assumed the same quantization axis for both electrodes. In the following $V^c$ and $V^s$ will be referred to as charge and spin voltages.

When charge and/or spin voltage is applied, the charge and spin currents, $J^c = J_\uparrow + J_\downarrow$ and $J^s = (J_\uparrow - J_\downarrow)/e$, respectively, can flow through the system. To calculate the currents one needs to determine the Green functions $G_\sigma^<$ and $G_\sigma^>$. We use here the Keldysh equation $G_\sigma^< = G_\sigma^r \Sigma_\sigma^< G_\sigma^a$ which allows us to relate the lesser Green function to the retarded $G^r$ and advanced $G^a$ ones. The self-energy $\Sigma^<$ is calculated with use of the Ng ansatz[37]: $\Sigma_\sigma^< = \Sigma_{0\sigma}^<(\Sigma_0^r - \Sigma_0^a)^{-1}(\Sigma_\sigma^r - \Sigma_\sigma^a)$ where $\Sigma_{0\sigma}^< = i(\Gamma_L f_{L\sigma} + \Gamma_R f_{R\sigma})$ is the lesser self-energy of the corresponding non-interacting system and $\Sigma_0^{r(a)} = \pm i(\Gamma_L + \Gamma_R)/2$. The appropriate self-energies for the interacting system are calculated according to the Dyson equation. In the case under consideration, the coupling parameters $\Gamma_\beta$ are independent of electron spin, and the lesser (greater) Green function can be expressed directly by the retarded $G_\sigma^r$ and advanced $G_\sigma^a$ Green functions. The latter can be easily determined by the equation of motion method.[12] To decouple the higher-order Green functions, we used the standard decoupling procedure proposed by Meir et al,[38] which allows to describe basic features of charge and spin transport in the Kondo regime.[36,38] The charge current $J_\sigma$ flowing in the spin channel σ can be then expressed in the well known form,

$$J_\sigma = \frac{ie}{\hbar}\int \frac{dE}{2\pi} \frac{\Gamma_L \Gamma_R}{\Gamma_L + \Gamma_R}[G_\sigma^r(E) - G_\sigma^a(E)][f_{L\sigma}(E) - f_{R\sigma}(E)]. \quad (4)$$

Assuming now the large-U limit, $U \to \infty$, one finds the Green function $G_\sigma^r(E)$ in the form $G_\sigma^r(E) = (1 - n_{-\sigma})(E - E_0 - \Sigma_0^r - \Sigma_{1\sigma}^r)^{-1}$, where $\Sigma_0^r = -i(\Gamma_L + \Gamma_R)/2$ while $\Sigma_{1\sigma}^{r} = \sum_{\beta} \int (dE/2\pi) \Gamma_{\beta} f_{\beta-\sigma} (E - E_{\sigma} + E_{-\sigma})^{-1}$ is a self-energy typical for the Kondo problem. The procedure applied here is described in details in Ref. 12, and is well justified at temperatures close to the Kondo temperature.

### B. Numerical results

Numerical calculations have been performed for the following parameters (measured in the units of $D/50$, with $D$ being the electron band width): the spin independent dot's level $E_0 = -0.35$, $\Gamma_L = \Gamma_R = 0.1$, and $k_B T = 0.001$. Such parameters are appropriate for transport in the Kondo regime.[12,13] At first we discuss spin and charge currents flowing through the system subject to a spin voltage only, $V^s \neq 0$ and $V^c = 0$. We will also distinguish between symmetric and asymmetric case (in our case the asymmetry will be introduced by an external magnetic field). Let us begin with the symmetric situation. In such a case, the charge current vanishes and only a spin current is nonzero. The mixed charge/spin differential conductance, defined as $G_{diff}^{cs} = dJ^c/dV^s$, vanishes then exactly, as shown by the solid line in Fig. 1b. In turn, the spin differential conductance is defined as $G_{diff}^{ss} = dJ^s/dV^s$, and is presented in Fig. 1a (solid curve). This conductance reveals the zero bias Kondo resonance which resembles the conductance behavior for a non-magnetic system subject to a charge bias. The Kondo anomaly in a spin-biased system also occurs in the density of states, given by $A_{\sigma}(E) = i(G_{\sigma}^{>} - G_{\sigma}^{<})$ and shown in Fig. 2a for several values of $V^s$. The Kondo anomaly appears at the Fermi energy ($E_F = 0$) for the unbiased system and splits into two components when the spin-bias is applied. Since the two spin channels are fully equivalent, the curves corresponding to different spin orientation (up and down) overlap in Fig. 2a.

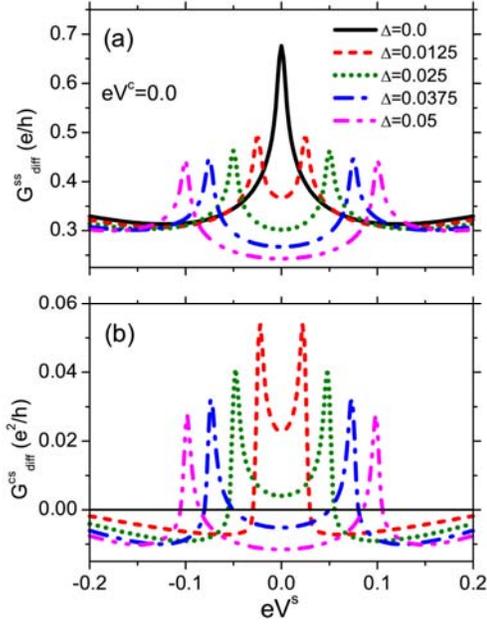

FIG. 1. Differential spin conductance $G_{diff}^{ss}$ (a) and mixed conductance $G_{diff}^{cs}$ (b) as a function of spin voltage for indicated values of the dot's level splitting $\Delta$, and in the absence of charge bias. The other parameters as described in the text.

Let us assume now the presence of an external magnetic field applied to the dot, which leads to a splitting of the dot energy level, $E_\sigma = E_0 - \hat{\sigma}\Delta$, where $\Delta$ represents magnitude of the splitting (proportional to the magnetic field). Since transport takes place through two different energy levels, the Kondo resonance in $G_{diff}^{ss}$ splits now into two components which move towards positive and negative values of $V^s$ with increasing $\Delta$ (see Fig. 1a). As follows from the figure, the Kondo anomaly is strongly suppressed for large values of the level splitting $\Delta$. The results obtained for $G_{diff}^{ss}$ are consistent with those found in Refs 28,29. As Katsura[28] used totally different method to describe the Kondo physics, the good consistency of the results shows that the present approach based on EOM can be found reliable. Presence of an external magnetic field leads to some asymmetry, so the two channels corresponding to spin-up and spin-down electrons are no longer equivalent and a small charge current can be observed. The corresponding mixed differential conductance $G_{diff}^{cs}$ shows two positive resonance peaks located on positive and negative sides of $V^s$ (Fig. 1b). It is interesting to note, that the conductance $G_{diff}^{cs}$ changes sign at some values of $V^s$, and becomes negative in relatively broad regions of $V^s$. The results shown in Fig. 1 correspond to positive values of $\Delta$. When the sign of $\Delta$ is changed, $\Delta \to -\Delta$, the spin conductance $G_{diff}^{ss}$ remains the same, whereas the mixed conductance $G_{diff}^{cs}$ changes sign and two negative resonance peaks can be observed. It should also be pointed out that the mixed conductance calculated for different positions of the dot level $E_0$ with respect to the Fermi energy in the leads is consistent with those obtained in Ref. 29. In particular, for $\Delta < 0$ and low-lying dot level, a flat maximum can be observed in the region of small spin voltages.

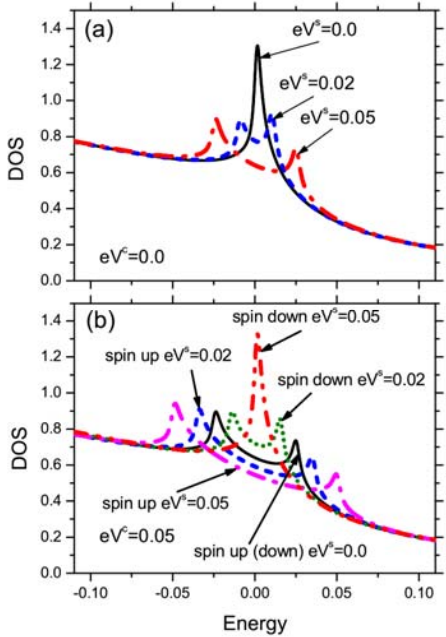

FIG. 2. Spin-dependent density of states (DOS) for indicated values of spin voltage in the absence of charge bias (a) and for $eV^c$=0.05 (b).

Consider now the situation when both charge and spin voltages are simultaneously applied. As before, let us consider first the interplay of these two types of voltages in the limit of spin-degenerate dot level, $\Delta = 0$. Spectral functions calculated for a fixed charge voltage, equal to $eV^c = 0.05$ and for three different values of spin voltage are presented in Fig. 2b. The two spin channels are equivalent for $V^s = 0$, and therefore the corresponding curves for spin-up and spin-down electrons overlap. The Kondo peak is then split into two components located at the Fermi levels of the two electrodes (shifted by $\pm V^c/2$ with respect to $E_F = 0$). When a spin voltage is additionally applied, the curves for spin-up and spin-down orientations behave differently. Splitting of the Kondo components for spin-up channel increases with increasing $V^s$, whereas splitting for spin-down electrons becomes reduced and is fully compensated when $eV^s = 0.05$. Then, only one well defined Kondo peak located at $E_F = 0$ can be observed for spin-down orientation.

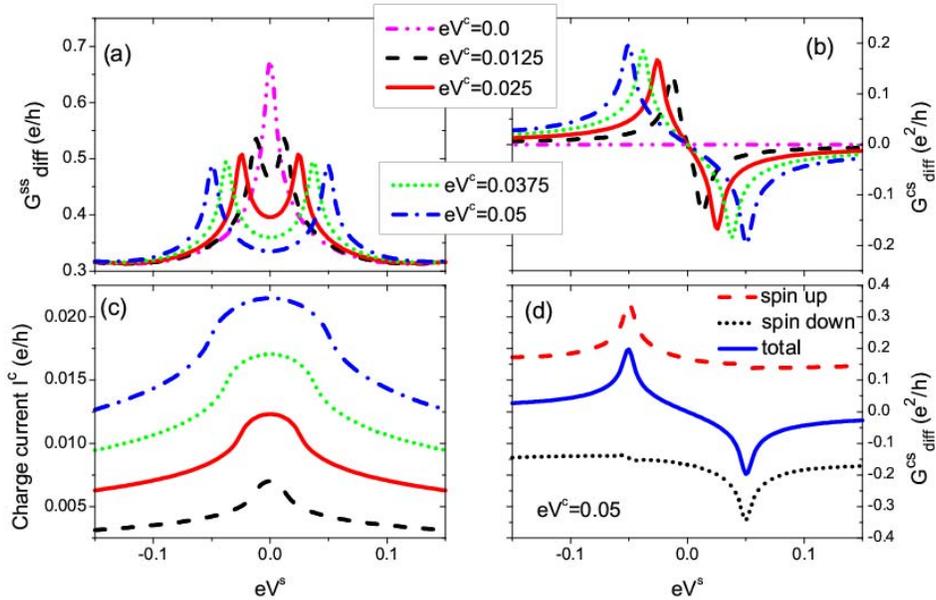

FIG. 3. Spin conductance (a), mixed conductance (b) and charge current (c) versus spin voltage for indicated values of charge voltage. Spin resolved mixed conductance for indicated charge voltage is shown in part (d).

The spin conductance $G_{diff}^{ss}$ as a function of the spin voltage is shown in Fig. 3a for several values of the charge voltage. Note, that the Kondo resonance becomes split into two components located at $\pm eV^c$ when a charge voltage $V^c$ is applied to the system, and the splitting increases with increasing $V^c$. Such a behavior is qualitatively similar to that for a non-magnetic Kondo system in an external magnetic field. However, in the case under consideration the splitting of the Kondo peak in the spin conductance is due to the charge voltage $V^c$ applied to the system in the absence of magnetic field. Significantly different behavior is revealed by the mixed conductance $G_{diff}^{cs}$, shown in Fig. 3b. First, $G_{diff}^{cs}$ changes sign when the spin voltage is reversed. Second, the corresponding curves have a shape characteristic of a resonance, with pronounced positive and negative peaks located at $eV^s = \mp eV^c$, symmetrically with respect to the zero spin voltage. Maximum absolute value of the conductance (at peaks) increases with increasing charge voltage. It is also interesting to point, that the conductance $G_{diff}^{cs}$ vanishes at $V^s = 0$, where however the charge current shows a well-defined maximum, as depicted in Fig. 3c. As one might expect, this maximum in current strongly increases with $V^c$.

Variation of $G_{diff}^{cs}$ can be understood when considering charge currents flowing in the spin-up and spin-down channels, separately. The spin-σ contribution to the mixed conductance, $G_{diff}^{cs\sigma} = dJ_\sigma / dV^s$, is shown in Fig. 3d. The conductance $G_{diff}^{cs\uparrow}$, associated with current flowing in the spin-up channel is positive and reveals a Kondo resonance located at $eV^s = -eV^c$, where the spin voltage compensates the charge voltage. Only a small step occurs then for $eV^s = eV^c$. However, the corresponding current in spin-down channel decreases with increasing spin bias, and the conductance $G_{diff}^{cs\downarrow}$ is negative and reveals the negative Kondo resonance located at $eV^s = eV^c$. When both contributions to $G_{diff}^{cs}$ are summed up, one obtains the mixed conductance as presented in Fig. 3b (and solid curve in Fig. 3d). It should be pointed out that Kondo peaks which appear in $G_{diff}^{cs}$ and $G_{diff}^{ss}$ at negative and positive spin voltages are due to electrons flowing in different spin channels (with spin up and down, respectively). Note, that in the absence of spin voltage ($V^s = 0$), currents in both spin channels flow in the same direction under a charge voltage $V^c$. The charge current reaches then a local maximum. When the positive spin voltage is applied, the current in spin-up channel increases, whereas the current in spin-down channel becomes considerably reduced and changes sign at some value of spin bias, which effectively leads to a decrease of charge current with increasing spin bias (see Fig. 3c).

The spin conductance as well as mixed conductance are displayed in Fig. 4 as a function of spin and charge voltages. The spin conductance is considerably enhanced in the region of small voltages, but the Kondo anomaly becomes split and strongly suppressed when the charge voltage or the spin voltage increases. On the other hand, the mixed conductance is close to zero in the zero bias (spin or charge) region, and a pronounced positive or negative resonance appears for large bias. The intensity of resonances increases with increasing spin and charge voltages.

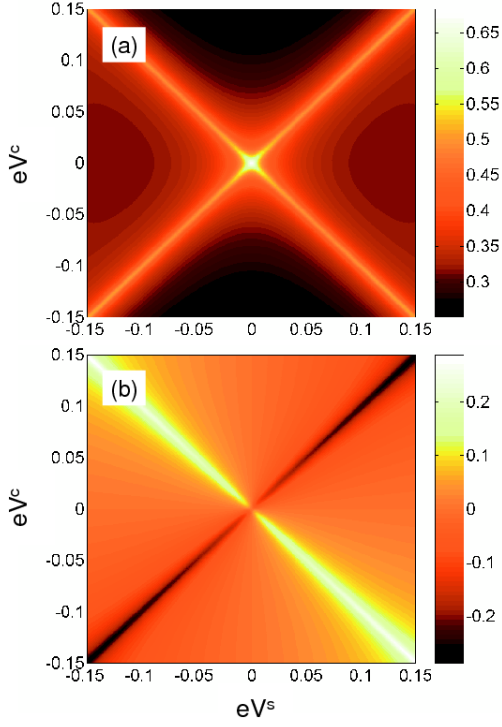

FIG. 4. The spin conductance (a) and the mixed conductance (b) versus the spin and charge voltages.

Now, we consider the effects due to an external magnetic field in the presence of both spin and charge voltages. The total mixed conductance $G_{diff}^{cs}$, as well as the contributions to $G_{diff}^{cs}$ from individual spin channels are presented in Fig. 5a for $eV^c = 0.05$ and $\Delta = 0.025$. As one can easily note, the resonance occurring in $G_{diff}^{cs\uparrow}$ at $eV^s = -eV^c$ in the absence of field is now split into two components which are shifted by $2\Delta$ towards higher and lower voltages, and appear at $eV^s = -eV^c - 2\Delta$ and $eV^s = -eV^c + 2\Delta$, respectively. Similar splitting also appears for $G_{diff}^{cs\downarrow}$, but now the conductance is negative.

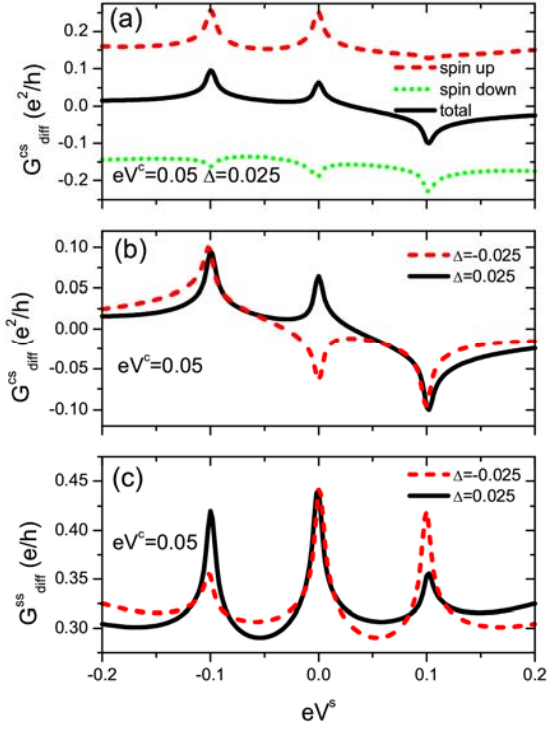

FIG. 5. Total mixed conductance and the corresponding contributions from individual spin channels (a), mixed conductance (b) and spin conductance (c) versus spin bias for indicated values of the dot level splitting and charge voltage.

The splitting of resonance in $G_{diff}^{cs\uparrow}$ and also that in $G_{diff}^{cs\downarrow}$ leads to a more complex behavior of the total mixed conductance $G_{diff}^{cs}$. For a chosen set of parameters, the two resonance peaks appear at $eV^s = -eV^c - 2\Delta$ and $eV^s = 0$. The intensity of the peak at $eV^s = 0$ is relatively small as the contributions from positive and negative resonances corresponding to opposite spin orientations partially compensate. Moreover, a well pronounced negative resonance appears at $eV^s = eV^c + 2\Delta$. The positive resonance peak located at $eV^s = 0$ can be easily converted into a negative resonance when the orientation of external magnetic field is reversed ($\Delta \to -\Delta$), as shown in Fig. 5b. Multiple Kondo peaks can be clearly seen in the spin conductance presented in Fig. 5c for $\Delta = \pm 0.025$ and $eV^c = 0.05$. For such a set of parameters, the peaks occurring at $eV^s = 0$ in $G_{diff}^{cs\uparrow}$ and $G_{diff}^{cs\downarrow}$ overlap leading to a resonant Kondo peak of a considerable intensity. Intensities of the side peaks depend on the sign of $\Delta$, and for positive $\Delta$ the peak appearing at negative spin voltages is considerably higher.

### III. THE INFLUENCE OF ELECTRON-PHONON COPUPLING IN THE DOT

Recent experimental data reveal an important role of vibrational degrees of freedom in electronic transport through quantum dots and molecules coupled to external leads.[39-47] Therefore, in this section we consider a quantum dot coupled to a phonon bath and analyze phonon-assisted transport under spin and charge bias applied to the system. To take into account the presence of vibrational modes, we extend the Hamiltonian $H$ by including the term $H_{ph} = \omega a^+ a$ which describes the phonon system. Here, $a^+$ and $a$ denote the creation and annihilation operators of a phonon with energy $\omega$, respectively. Moreover, the electron-phonon interaction term is introduced into the dot Hamiltonian which now takes the following form:

$$H_D = \sum_\sigma E_0 d_\sigma^+ d_\sigma + U d_\uparrow^+ d_\uparrow d_\downarrow^+ d_\downarrow + \lambda(a^+ + a)\sum_\sigma d_\sigma^+ d_\sigma, \qquad (5)$$

with $\lambda$ being the electron-phonon coupling parameter.

To calculate spin-dependent current $J_\sigma$ according to the formula (3), one should determine the Green functions $G_\sigma^{<(>)}$ in the presence of phonons. Here, we proceed the approach described in details in our earlier paper,[48] so only very brief description is given. A standard procedure based on the Lang-Firsov transformation, and discussed in many relevant publications (see e.g. Refs 48-51), is applied to eliminate electron-phonon term from the dot Hamiltonian (5). The two sub-systems become decoupled when the phonon operators in the transformed Hamiltonian are replaced by their expectation values in thermal equilibrium. The QD is then described by Hamiltonian $\tilde{H}_D$ of the form given by Eq. (2), but with renormalized energy level, $E_0 \rightarrow \tilde{E}_0 = E_0 - g\omega$, and renormalized correlation parameter, $U \rightarrow \tilde{U} = U - 2g\omega$, where $g = (\lambda/\omega)^2$. The parameters that describe coupling of the dot and leads are also renormalized, $\Gamma_\beta \rightarrow \tilde{\Gamma}_\beta = \Gamma_\beta \exp[-g(2N_{ph}+1)]$, with $N_{ph}$ denoting the equilibrium phonon population. The Fourier transform of the lesser (greater) Green function can be written in the form $G_\sigma^{<(>)}(E) = \sum_{n=-\infty}^{\infty} L_n \tilde{G}_\sigma^{<(>)}(E \pm n\omega)$, where the upper (lower) sign refers to $G_\sigma^<$ ($G_\sigma^>$), and $L_n = e^{-g(2N_{ph}+1)} e^{n\omega/2k_B T} I_n(2g\sqrt{N_{ph}(N_{ph}+1)})$ with $I_n(z)$ being the n-th Bessel function of complex argument. As electron and phonon sub-systems are decoupled, and the transformed Hamiltonian has the form similar to the one discussed in the previous section with the dot term being typical Anderson Hamiltonian, the Green functions $\tilde{G}_\sigma^{<(>)}$ are calculated using the formalism described in the preceding section.

The results obtained for $\omega = 0.1$ and several values of the coupling parameter $g$ are presented in Fig. 6, where the differential spin conductance is shown as a function of spin voltage. When the transport voltage $V^c$ is equal to zero, the spin conductance $G_{diff}^{ss}$ shows a typical Kondo behavior with a resonance peak in the zero spin voltage regime. Due to the electron-phonon coupling, the satellites of the Kondo peak appear on both sides of the main Kondo resonance at the energies $\pm\omega$ for the first-order satellites and $\pm 2\omega$ for the second-order ones. The satellites, particularly the second order ones, can be clearly visible for higher values of $g$ (Fig. 6a). This result is in a close analogy to the one obtained for the charge conductance $G_{diff}^{cc} = dJ^c/dV^c$ of a typical Kondo system subject to the transport voltage.[48,50-51] On the other hand, a different behavior can be observed when the charge voltage $V^c$ is applied apart from the spin bias (Fig. 6b). The main Kondo resonance becomes then split into two components which appear at energies $\pm eV^c$. The Kondo satellites due to coupling to the phonon bath occur at the distance $\omega$ from the main peaks. The second satellites at the distance $2\omega$ are also well seen. However, no phonon features can be observed at energies $\pm n\omega$ ($n$=1,2..).

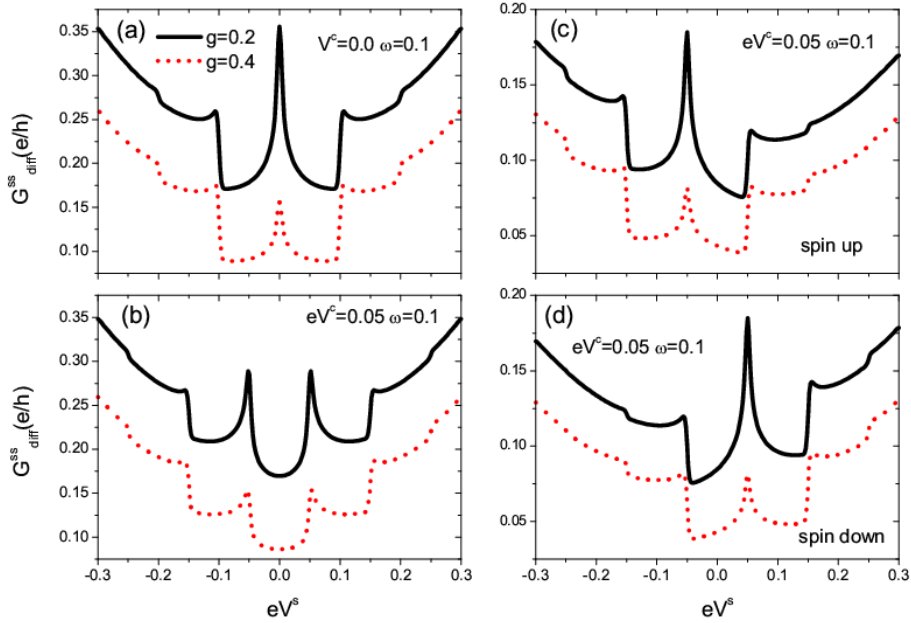

FIG. 6. The spin conductance as a function of spin voltage for indicated values of $g$ and in the absence of charge voltage, $eV^c = 0$ (a), and with charge voltage applied, $eV^c = 0.05$ (b). Contributions to the spin conductance from spin-up (c) and spin-down (d) channels.

Contributions from individual spin channels to the differential spin conductance, $G_{diff}^{ss} = G_{diff}^{ss\uparrow} + G_{diff}^{ss\downarrow}$, are shown in Figs 6c and 6d. The spin-up contribution, $G_{diff}^{ss\uparrow}$, is positive and shows the main Kondo resonance at the energy $-eV^c$ as well as satellite peaks at distances $\omega$ and $2\omega$ on both sides of the main resonance. The curves are not symmetric with respect to the spin voltage reversal as a nonzero charge voltage $V^c$ is applied. Note, that the curves are not only shifted towards negative spin voltages by a value $eV^c$, but an additional asymmetry in the intensities of phonon satellite peaks appears. The spin-down contribution, $G_{diff}^{ss\downarrow}$, is also positive and exhibits resonance located at $eV^c$ as well as an additional structure due to coupling to phonons. The curves for spin-down orientation are shifted by $eV^c$ towards positive values of spin voltages. Since the curves corresponding to spin-up and spin-down contributions are shifted in opposite directions, the two main resonances appear in the total spin conductance $G_{diff}^{ss}$. For the chosen set of parameters, however, the original Kondo peak for one spin orientation overlaps with a satellite peak corresponding to the opposite spin, and therefore no additional features can be seen in the total spin conductance.

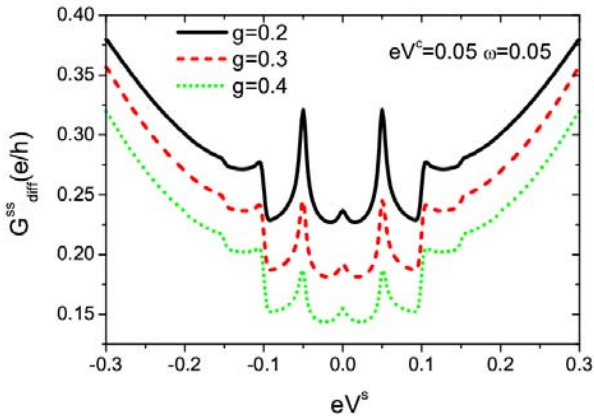

FIG. 7. The spin conductance as a function of spin voltage for $eV^c = 0.05$, $\omega = 0.05$ and indicated values of the parameter *g*.

These additional features can be observed for a different set of parameters, e.g. for $\omega = 0.05$, as depicted in Fig. 7. Now, apart from two well pronounced Kondo peaks at energies $\pm eV^c$ and their satellites which occur for $\pm(eV^c + \omega)$ and $\pm(eV^c + 2\omega)$, a small cusp between the main Kondo peaks appears for $V^s = 0$. It corresponds to the overlapping Kondo satellites coming from the spin-up and spin-down contributions. In general, for a given system coupled to the vibrational mode of energy $\omega$ one can easily tune the charge voltage to change structure of the conductance curve by changing positions of the Kondo peaks and their satellites, as well as the resonance intensities when certain peaks overlap.

## IV. SUMMARY AND CONCLUSIONS

We have investigated charge and spin transport through a quantum dot in the Kondo regime, with particular emphasis put on the interplay of spin and charge voltages. In general, the spin conductance $G_{diff}^{ss}$ as a function of the spin voltage exhibits features characteristic of a standard Kondo system. A well-pronounced Kondo anomaly occurs in the zero spin bias regime. The resonance peak becomes split into two components by external magnetic field. Very similar behavior of the Kondo anomaly is observed when the charge voltage is applied, which also leads to a splitting of the peak and the two components appear at $eV^s = \pm eV^c$. Both magnetic field and charge voltage significantly suppress the Kondo effect. However, some novel features resulting in the multiple Kondo resonances have been revealed due to the interplay of charge voltage and external magnetic field. The number of peaks and their intensities can be easily changed by tuning the voltage $V^c$ or the field intensity. The zero bias Kondo anomaly, split under the influence of magnetic field, can be restored when the charge voltage is applied to the system.

The mixed charge/spin conductance, $G_{diff}^{cs}$, shows a different behavior. In the zero spin bias limit, $G_{diff}^{cs}$ is equal to zero despite of the presence of charge voltage. At the same time, the charge current shows the round maximum at $V^s = 0$ and a non-vanishing charge voltage. Moreover, new features have been found in the spin voltage dependence of $G_{diff}^{cs}$. More specifically, a positive and negative resonances due to the charge voltage can be observed.

This work was supported by funds of the Polish Ministry of Science and Higher Education as a research project in years 2006-2009.